\documentclass[twocolumn]{aastex631}

\usepackage{xcolor}

\hypersetup{linkcolor=red,citecolor=blue,filecolor=cyan,urlcolor=black}
\newcommand{\grizli}{\texttt{Grizli}}

\shorttitle{First spectroscopic confirmation of low-mass quiescent galaxies at $z>2$}
\shortauthors{Marchesini et al.}

\graphicspath{{./}{figures/}}

\begin{document}

\title{Early results from GLASS-JWST. IX: First spectroscopic confirmation of low-mass quiescent galaxies at $z>2$ with NIRISS}

\correspondingauthor{Danilo Marchesini}
\email{danilo.marchesini@tufts.edu}

\author[0000-0001-9002-3502]{Danilo Marchesini}
\affiliation{Physics and Astronomy Department, Tufts University, 574 Boston Avenue, Medford, MA 02155, USA}

\author[0000-0003-2680-005X]{Gabriel Brammer}
\affiliation{Cosmic Dawn Center (DAWN), Denmark}
\affiliation{Niels Bohr Institute, University of Copenhagen, Jagtvej 128, DK-2200 Copenhagen N, Denmark}

\author[0000-0002-8512-1404]{Takahiro Morishita}
\affiliation{IPAC, California Institute of Technology, MC 314-6, 1200 E. California Boulevard, Pasadena, CA 91125, USA}

\author[0000-0003-1383-9414]{Pietro Bergamini}
\affiliation{Dipartimento di Fisica, Università degli Studi di Milano, Via Celoria 16, I-20133 Milano, Italy}
\affiliation{INAF - OAS, Osservatorio di Astrofisica e Scienza dello Spazio di Bologna, via Gobetti 93/3, I-40129 Bologna, Italy}

\author[0000-0002-9373-3865]{Xin Wang}
\affil{Infrared Processing and Analysis Center, Caltech, 1200 E. California Blvd., Pasadena, CA 91125, USA}

\author[0000-0001-5984-0395]{Marusa Bradac}
\affiliation{University of Ljubljana, Department of Mathematics and Physics, Jadranska ulica 19, SI-1000 Ljubljana, Slovenia}
\affiliation{Department of Physics and Astronomy, University of California Davis, 1 Shields Avenue, Davis, CA 95616, USA}

\author[0000-0002-4140-1367]{Guido Roberts-Borsani}
\affiliation{Department of Physics and Astronomy, University of California, Los Angeles, 430 Portola Plaza, Los Angeles, CA 90095, USA}

\author[0000-0002-6338-7295]{Victoria Strait}
\affiliation{Cosmic Dawn Center (DAWN), Denmark}
\affiliation{Niels Bohr Institute, University of Copenhagen, Jagtvej 128, DK-2200 Copenhagen N, Denmark}

\author[0000-0002-8460-0390]{Tommaso Treu}
\affiliation{Department of Physics and Astronomy, University of California, Los Angeles, 430 Portola Plaza, Los Angeles, CA 90095, USA}

\author[0000-0003-3820-2823]{Adriano Fontana}
\affiliation{INAF Osservatorio Astronomico di Roma, Via Frascati 33, 00078 Monteporzio Catone, Rome, Italy}

\author[0000-0001-5860-3419]{Tucker Jones}
\affiliation{Department of Physics and Astronomy, University of California Davis, 1 Shields Avenue, Davis, CA 95616, USA}

\author[0000-0002-9334-8705]{Paola Santini}
\affiliation{INAF Osservatorio Astronomico di Roma, Via Frascati 33, 00078 Monte Porzio Catone, Rome, Italy}

\author[0000-0003-0980-1499]{Benedetta Vulcani}
\affiliation{INAF Osservatorio Astronomico di Padova, vicolo dell'Osservatorio 5, 35122 Padova, Italy}

\author[0000-0003-3108-9039]{Ana~Acebron}
\affiliation{Dipartimento di Fisica, Università degli Studi di Milano, Via Celoria 16, I-20133 Milano, Italy}
\affiliation{INAF - IASF Milano, via A. Corti 12, I-20133 Milano, Italy}

\author[0000-0003-2536-1614]{Antonello Calabr\`o}
\affiliation{INAF Osservatorio Astronomico di Roma, Via Frascati 33, 00078 Monteporzio Catone, Rome, Italy}

\author[0000-0001-9875-8263]{Marco~Castellano}
\affiliation{INAF Osservatorio Astronomico di Roma, Via Frascati 33, 00078 Monte Porzio Catone, Rome, Italy}

\author[0000-0002-3254-9044]{Karl Glazebrook}
\affiliation{Centre for Astrophysics and Supercomputing, Swinburne University of Technology, PO Box 218, Hawthorn, VIC 3122, Australia}

\author[0000-0002-5926-7143]{Claudio Grillo}
\affiliation{Dipartimento di Fisica, Università degli Studi di Milano, Via Celoria 16, I-20133 Milano, Italy}
\affiliation{INAF - IASF Milano, via A. Corti 12, I-20133 Milano, Italy}

\author[0000-0001-9261-7849]{Amata Mercurio}
\affiliation{INAF -- Osservatorio Astronomico di Capodimonte, Via Moiariello 16, I-80131 Napoli, Italy}

\author[0000-0003-2804-0648 ]{Themiya Nanayakkara}
\affiliation{Centre for Astrophysics and Supercomputing, Swinburne University of Technology, PO Box 218, Hawthorn, VIC 3122, Australia}

\author[0000-0002-6813-0632]{Piero Rosati}
\affiliation{Dipartimento di Fisica e Scienze della Terra, Università degli Studi di Ferrara, Via Saragat 1, I-44122 Ferrara, Italy}
\affiliation{INAF - OAS, Osservatorio di Astrofisica e Scienza dello Spazio di Bologna, via Gobetti 93/3, I-40129 Bologna, Italy}

\author[0000-0002-7907-2634]{Chanita Tubthong}
\affiliation{Physics and Astronomy Department, Tufts University, 574 Boston Avenue, Medford, MA 02155, USA}

\author[0000-0002-5057-135X]{Eros~Vanzella}
\affiliation{INAF -- OAS, Osservatorio di Astrofisica e Scienza dello Spazio di Bologna, via Gobetti 93/3, I-40129 Bologna, Italy}

\begin{abstract}
How passive galaxies form, and the physical mechanisms which prevent star formation over long timescales, are some of the most outstanding questions in understanding galaxy evolution. The properties of quiescent galaxies over cosmic time provide crucial information to identify the quenching mechanisms. Passive galaxies have been confirmed and studied out to $z\sim4$, but all of these studies have been limited to massive systems (mostly with $\log{(M_{\rm star}/M_{\odot})}>10.8$). Using James Webb Space Telescope (JWST) NIRISS grism slitless spectroscopic data from the GLASS JWST ERS program, we present spectroscopic confirmation of two quiescent galaxies at $z_{\rm spec}=2.650^{+0.004}_{-0.006}$ and $z_{\rm spec}=2.433^{+0.032}_{-0.016}$ (3$\sigma$ errors) with stellar masses of $\log{(M_{\rm star}/M_{\odot})}=10.53^{+0.18}_{-0.06}$ and $\log{(M_{\rm star}/M_{\odot})}=9.93^{+0.06}_{-0.07}$ (corrected for magnification factors of $\mu=2.0$ and $\mu=2.1$, respectively). The latter represents the first spectroscopic confirmation of the existence of low-mass quiescent galaxies at cosmic noon, showcasing the power of JWST to identify and characterize this enigmatic population.  
\end{abstract}


%\keywords{Galaxies (573) --- High-redshift galaxies(734) --- Quenched galaxies(2016) --- Galaxy spectroscopy(2171) --- Galaxy evolution(594) --- Galaxy quenching(2040)}



\section{Introduction} \label{sec:intro}
Two of the most outstanding questions in understanding galaxy evolution are how quiescent galaxies form and what mechanisms are responsible for the halting of the star formation activity. Several physical processes and theoretical explanations have been suggested to first stop star formation in a galaxy and then maintain it quiescent over its lifetime (see \citealt{ManBelli2018} for a brief review of quenching mechanisms). The properties of quiescent galaxies as a function of cosmic time (e.g., stellar mass function and number density, star-formation history, stellar age, metallicity, [$\alpha$/Fe] abundance, size and morphology, kinematics, and environment) provide crucial information to identify the quenching mechanisms and their relative importance at different times in the cosmic history.

Over the last decade, incredibly dedicated effort and investment on ground-based 8-10m telescopes and the Hubble Space Telescope (HST) repeatedly pushed the spectroscopic confirmation of the existence of quiescent galaxies to $z\sim4$, when the Universe was only 1.5 Gyr old, progressively challenging theoretical models of galaxy formation and evolution (\citealt{Gobat2012}; \citealt{Marsan2015}; \citealt{Glazebrook2017}; \citealt{KadoFong2017}; \citealt{Forrest2020a}). Multi-object deep near-infrared (NIR) (i.e., rest-frame optical) spectroscopy of $2<z<4$ quiescent galaxies has allowed for the measurements of their stellar populations and star-formation histories (\citealt{Schreiber2018}; \citealt{Belli2019}; \citealt{DEugenio2020, DEugenio2021}; \citealt{EstradaCarpenter2020}; \citealt{Forrest2020a, Forrest2020b}; \citealt{Saracco2020}; \citealt{Valentino2020}), kinematics, sizes, and morphologies (\citealt{vandeSande2013}; \citealt{Belli2017}; \citealt{Hill2016}; \citealt{Newman2018}; \citealt{Tanaka2019}; \citealt{EstradaCarpenter2020}; \citealt{Saracco2020}; \citealt{Stockmann2021}; \citealt{Esdaile2021}; \citealt{Forrest2022}), metallicity (\citealt{Morishita2019}; \citealt{Saracco2020}), [$\alpha$/Fe] abundances \citep{Kriek2016}, and environment \citep{McConachie2022}. 

However, all of these studies have been limited to the most massive quiescent galaxies (mostly $\log{(M_{\rm star}/M_{\odot})}>11$, with only a small fraction with $10.5<\log{(M_{\rm star}/M_{\odot})}<11$). No low-mass (i.e., $\log{(M_{\rm star}/M_{\odot})}<10.5$, hereafter; corresponding to $\sim$1/3 of the characteristic stellar mass of the galaxy stellar mass function, e.g., \citealt{Muzzin2013}) quiescent galaxy at $z>2$ has been spectroscopically confirmed yet, preventing the exploration of quenching mechanisms to the low stellar mass regime at cosmic noon. For example, while some form of feedback generally attributed to active galactic nuclei (AGN) is supposed for the quenching of galaxies at the massive end, it is unclear whether AGN feedback can also operate in low-mass galaxies. Furthermore, halo quenching is not expected to operate in dark matter halos below $10^{12}$~M$_{\odot}$ at $z\approx2$ \citep{Dekel2008}, corresponding to roughly $M_{\rm star}\sim10^{10.5}$~M$_{\odot}$ \citep{Behroozi2013}. Indeed, multi-wavelength photometric surveys have been able to push photometric measurements of the stellar mass function of quiescent galaxies down to $\log{(M_{\rm star}/M_{\odot})}\sim9.5-10$ at $z\approx2-3$, finding a sharp decline in the number density of quiescent galaxies with $\log{(M_{\rm star}/M_{\odot})} \lesssim 10.5$ (e.g., \citealt{Tomczak2014}), although spectroscopic redshift measurements are needed to confirm these findings. 

The advent of the James Webb Space Telescope (JWST) with its amazing set of instruments finally enables us to definitively identify low-mass quiescent galaxies at $z>2$ and to spectroscopically characterize their properties for the first time.

In this Letter, we present spectroscopic confirmation of two quiescent galaxies at $z>2$, including the first spectroscopic confirmation of a low-mass quiescent galaxy at $z>2$, using JWST NIRISS grism slitless spectroscopic data obtained from the GLASS JWST Early Release Science (ERS-1324; PI Treu) program. GLASS-JWST is obtaining NIRISS (\citealt{Doyon2012, Willott2022}) and NIRSpec (\citealt{Jakobsen2022}) spectroscopy in the center of the A2744 cluster field, while obtaining NIRCAM (\citealt{Rieke2005}) images of two parallel fields. GLASS-JWST consists of the deepest extragalactic data amongst the ERS programs. Details can be found in the survey paper \citep{Treu2022}. 

This Letter is organized as follows. In Section~\ref{sec:sample} we introduce the selection of the low-mass quiescent galaxy candidates from multi-wavelength photometric catalogs in the Abell2744 cluster field. In Section~\ref{sec:analysis} we present the JWST-NIRISS grism spectra of the low-mass quiescent galaxies, their spectroscopic redshifts, robust stellar masses obtained from the modeling of the photometry, and gravitational lensing magnification corrections. We conclude in Section~\ref{sec:conclusions}. Magnitudes are given in the AB system and a standard cosmology with $\Omega_{\rm m}=0.3, \Omega_{\rm \Lambda}=0.7$ and $h=0.7$ is assumed when necessary, along with a \citet{Chabrier2003} initial mass function (IMF). 

\begin{figure}
\center
 \includegraphics[width=3.2in]{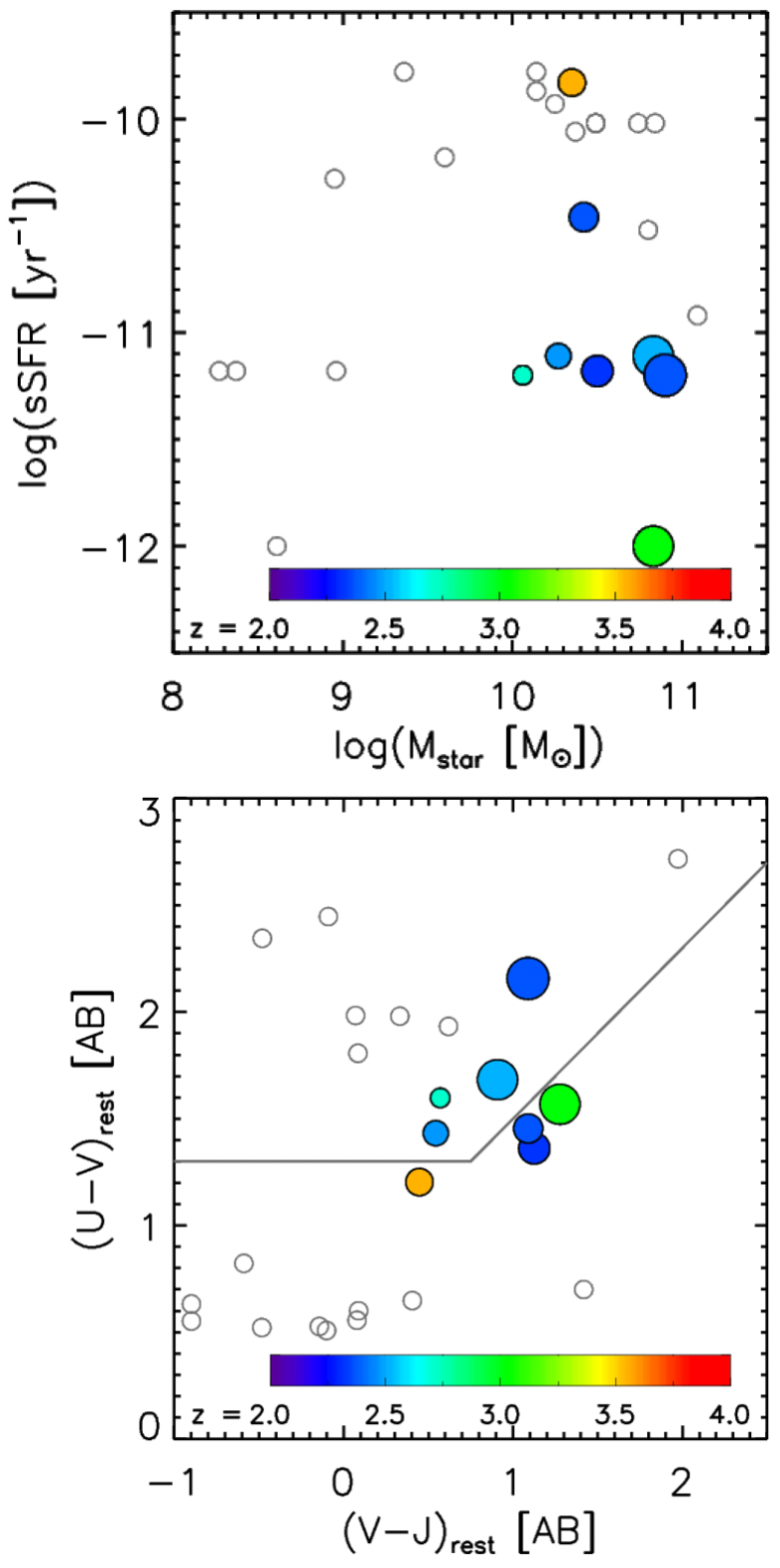}
 \caption{{\bf Top:} The sSFR vs $M_{\rm star}$ diagram of galaxies selected as explained in \S~\ref{sec:sample}. The candidates of low-mass quiescent galaxies at $z>2$ are shown color-coded as a function of redshift; larger colored filled circles indicate larger stellar masses.  {\bf Bottom:} UVJ diagram of the same galaxies plotted in the top panel, with the quiescent wedge as defined in \citet{Martis2016} plotted in gray. {\it Eight robust candidates of low-mass quiescent galaxies at $z>2$ are identified.}}
 \label{fig:SFR_Mstar}
\end{figure}

\section{Sample of low-mass quiescent galaxies} \label{sec:sample}

Candidate low-mass quiescent galaxies were selected from the HFF-DeepSpace catalogs in the Abell~2744 cluster (A2744-clu, hereafter) pointing constructed by \citet{Shipley2018}, and cross validated using the HFF-ASTRODEEP catalogs \citep{Castellano2016,Merlin2016}. 

The NIR-selected HFF-DeepSpace A2744-clu photometric catalog covers an area of $\sim$5.4 arcmin$^{2}$ with complete coverage in 14 bands from the HST-UVIS F275W to the {\it Spitzer}-IRAC 8~$\mu$m down to a 75\% detection completeness of $\sim$28.2 mag in the F160W band. EAZY \citep{Brammer2008} was used to derive photometric redshifts ($z_{\rm phot}$) and rest-frame colors. FAST \citep{Kriek2009} was used to derive stellar population properties adopting \citet{Bruzual2003} stellar population synthesis models, a delayed exponentially declining star-formation history (SFH)\footnote{SFR $\propto t~e^{-t/tau}$}, solar metallicity, and a \citet{Calzetti2000} dust attenuation law with $0<A_{\rm V}{\rm [mag]}<6$. See \citet{Shipley2018} for a detailed description of the HFF-DeepSpace catalogs and high-level science data.

\begin{figure}
\center
 \includegraphics[width=3.4in]{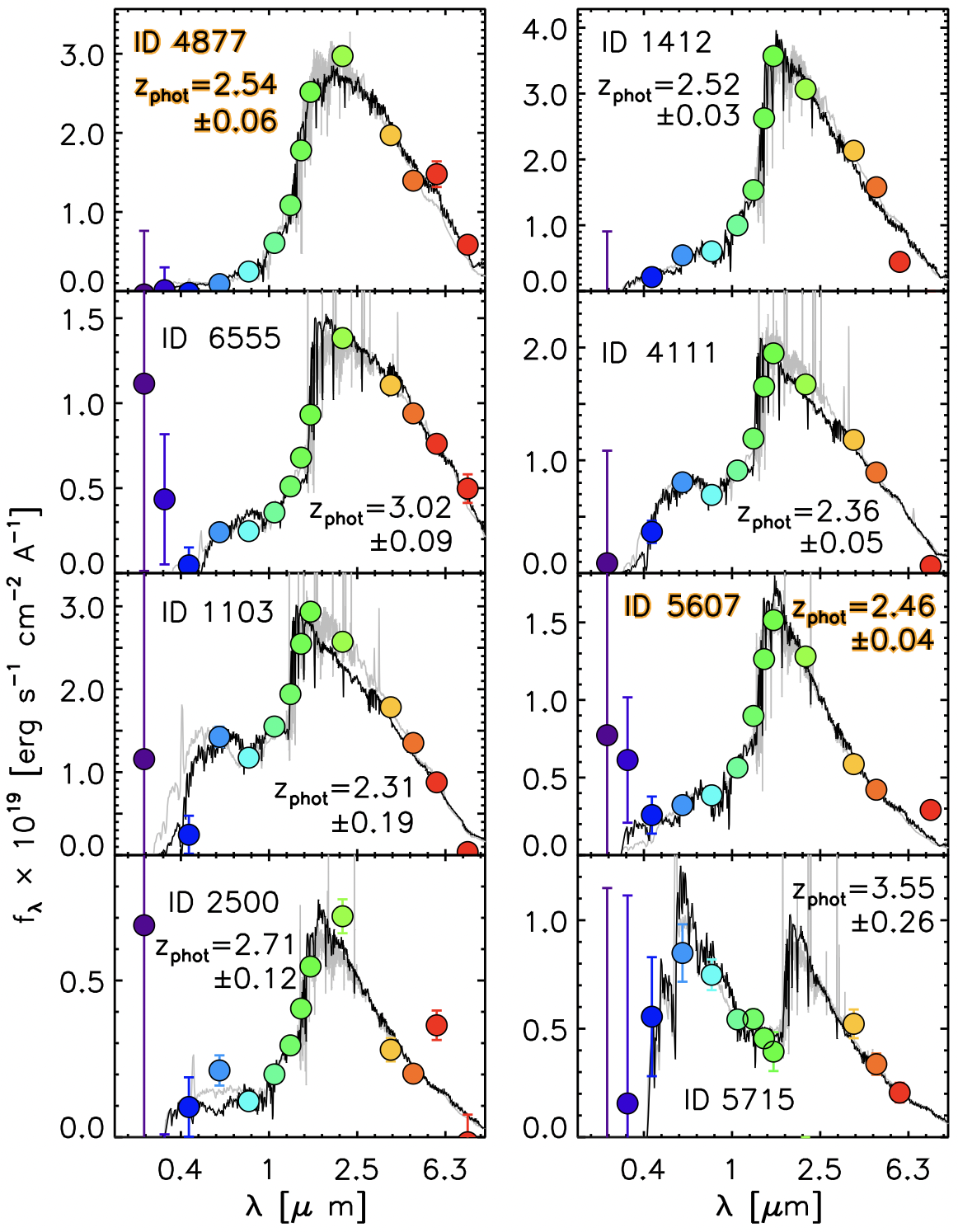}
 \caption{SEDs of the selected candidates of low-mass quiescent galaxies at $z>2$. Filled colored circles represent the observed HFF-DeepSpace photometry; the gray and black curves show the best-fit EAZY and FAST models, respectively. The photometric redshifts and the corresponding 1-$\sigma$ errors are listed. The two galaxies (IDs=4877 and 5607) with NIRISS spectroscopy presented in this Letter are highlighted in orange.}
 \label{fig:sedQGs}
\end{figure}

\begin{table*}
    \centering
    \begin{tabular}{ccccccccc}
        \hline
        ID & RA & DEC & H$_{\rm 160}$ & $z_{\rm phot}$ & $\rm \log{(\mu \times M_{star})}$ & $\rm \log{(sSFR)}$ & $\mu$ & $\rm \log{(M^{\rm corr}_{star})}$   \\
        
        [HFF-DS] & [deg] & [deg] & [AB] & & [M$_{\odot}$] & [yr$^{-1}$] & & [M$_{\odot}$] \\
        \hline
        4877$^{\dagger}$ & 3.605219 & -30.39540 & 23.15 & 2.54$\pm$0.06 & 10.97 & -11.20 & 2.0$^{+0.1}_{-0.1}$ & 10.67 \\
        1412 & 3.575537 & -30.41321 & 22.77 & 2.52$\pm$0.03 & 10.83 & -11.11 & 2.1$^{+0.1}_{-0.1}$ & 10.52 \\
        6555 & 3.582519 & -30.38545 & 24.23 & 3.02$\pm$0.09 & 10.83 & -12.00 & 4.0$^{+0.1}_{-0.1}$ & 10.23 \\
        4111 & 3.568355 & -30.39953 & 23.43 & 2.36$\pm$0.05 & 10.42 & -10.46 & 2.3$^{+0.1}_{-0.1}$ & 10.06 \\
        1103 & 3.585363 & -30.41533 & 22.98 & 2.31$\pm$0.19 & 10.50 & -11.18 & 2.8$^{+0.1}_{-0.1}$ & 10.06 \\
        5607$^{\dagger}$ & 3.601594 & -30.39150 & 23.70 & 2.46$\pm$0.04 & 10.24 & -11.11 & 2.1$^{+0.1}_{-0.1}$ & 9.92 \\
        2500 & 3.572230 & -30.40686 & 24.81 & 2.71$\pm$0.12 & 10.06 & -11.20 & 2.2$^{+0.1}_{-0.1}$ & 9.73 \\
        5715 & 3.586502 & -30.39048 & 25.16 & 3.55$\pm$0.26 & 10.35 & -9.83 & 9.1$^{+1.1}_{-0.6}$ & 9.39 \\
        \hline 
    \end{tabular}
    \caption{Sample of selected candidates of low-mass quiescent galaxies at $z>2$ and their properties (from \citealt{Shipley2018}) derived adopting the listed photometric redshifts $z_{\rm phot}$ (with 1$\sigma$ uncertainties). $\rm \log{(M^{\rm corr}_{star})}$ lists the stellar mass after correcting for the listed gravitational lensing magnification factors $\mu$ from \citet{Bergamini2022}; the quoted uncertainties on $\mu$ are 1-$\sigma$ statistical errors. $^{\dagger}$The two galaxies with NIRISS spectroscopy presented in this Letter.}
    \label{tab:QGs}
\end{table*}

First, we selected all galaxies with $m_{\rm F160W}<26.5$, photometric redshift $z_{\rm phot}>2$, stellar mass $M_{\rm star}>10^{8}$~M$_{\odot}$, and specific star-formation rate (sSFR) $\log{(sSFR[yr^{-1}])}<-9.7$, and use\_phot=1\footnote{The flag use\_phot$=$1 excludes stars, sources close to a bright star, sources with $S/N<3$ in the F160W band, catastrophic photometric redshift fits, and catastrophic stellar population fits; see \citet{Shipley2018}.}. We then visually inspected postage stamps of the HST images, detection image, and segmentation map of all resulting objects using the HFFexplorer\footnote{http://cosmos.phy.tufts.edu/$\sim$danilo/HFF/HFFexplorer/} to remove any spurious or problematic objects (e.g., galaxies contaminated by stellar diffraction spikes, objects detected too close to subtracted bright cluster galaxies, very broad/unconstrained photometric redshift probability functions, blended objects, and potentially merging systems). This resulted in 25 galaxies, shown in the sSFR--$M_{\rm star}$ diagram (Fig.~\ref{fig:SFR_Mstar}, top panel) and in the rest-frame $U-V$ vs $V-J$ diagram (Fig.~\ref{fig:SFR_Mstar}, bottom panel; UVJ diagram, hereafter).

Candidate quiescent galaxies were finally selected based on their positions within (or close to the boundary of) the quiescent wedge in the UVJ diagram and from visual inspection of their spectral energy distributions (SEDs), looking for strong rest-frame optical Balmer/4000~\AA~ breaks. A few additional objects with poorly sampled SEDs around the rest-frame optical breaks were removed from the initial sample. This resulted in the identification of eight low-mass quiescent galaxies shown as filled colored circles in Figure~\ref{fig:SFR_Mstar}. Their SEDs are shown in Figure~\ref{fig:sedQGs}, while Table~\ref{tab:QGs} lists their photometrically derived properties. 

Several works (e.g., \citealt{Marsan2015, Merlin2018, Merlin2019, Schreiber2018, Carnall2020}) have shown that the standard selection of quiescent galaxies in the UVJ diagram (in particular the $U-V>1.3$ criterion) is potentially biased against unobscured recently quenched (i.e., younger than a few hundred Myr) post-starburst galaxies, which still have significant emission in the UV due to massive bright stars not yet evolved off the main sequence (see, e.g., ID=5715 in Fig.~\ref{fig:sedQGs}). Here we apply a conservative approach, selecting unambiguous evolved quiescent galaxies, and we postpone a systematic and complete search for quiescent distant galaxies to future studies.
 
\begin{figure*}
\center
 \includegraphics[width=\textwidth]{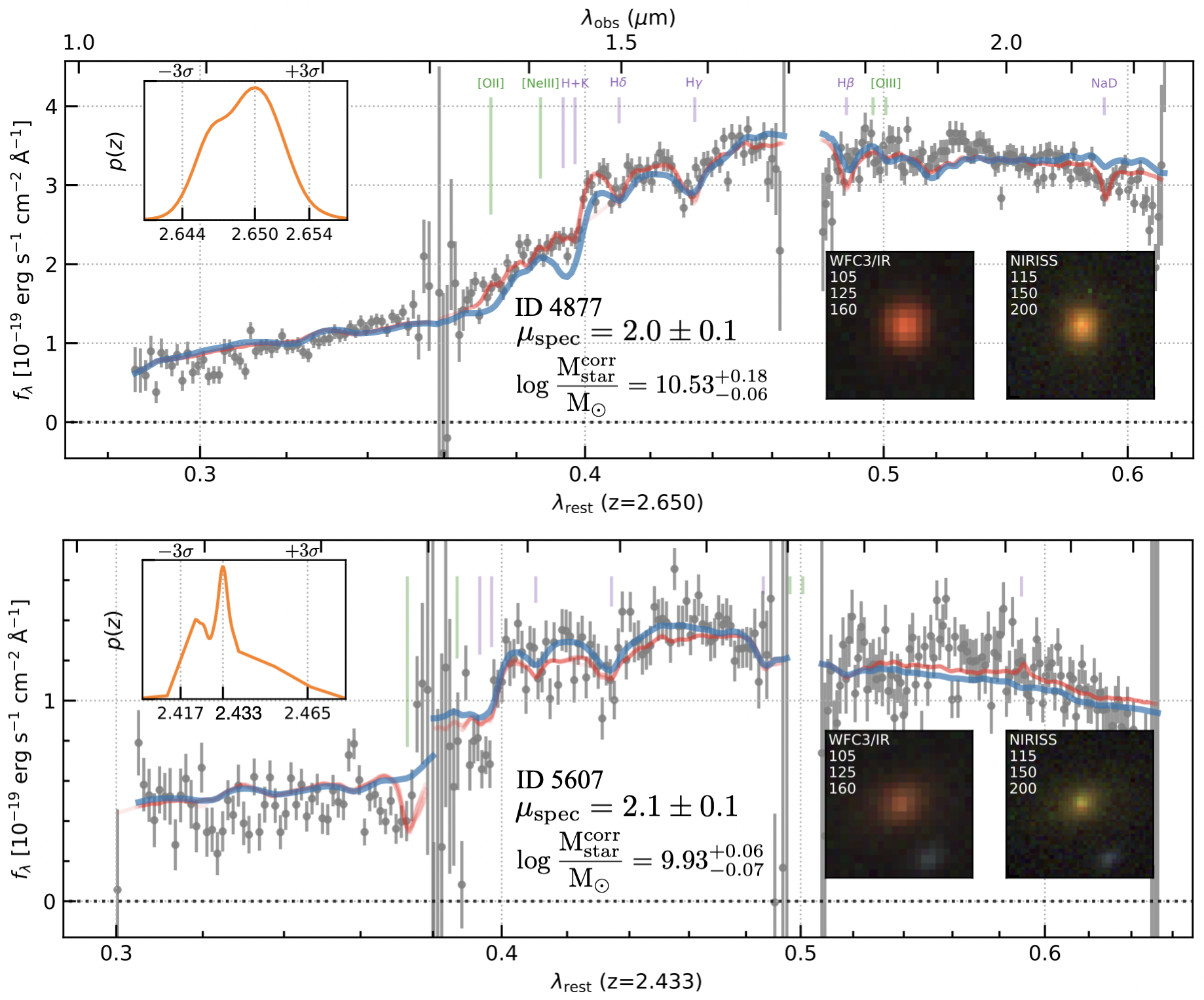}
 \caption{{\bf Top:} NIRISS F115W+F150W+F200W spectrum of HFF-DS 4877 (gray symbols) with \grizli\ best-fit model (red solid curve) and FAST best-fit model (blue solid curve) overplotted. the position of major spectral features (e.g., [OII], CaII HK, H$\delta$, H$\gamma$, H$\beta$, [OIII], and NaD) are highlighted at the best-fit \grizli\ redshift. The gravitational lensing magnification factor and stellar mass, corrected for lensing magnification, derived assuming the spectroscopic redshift are listed. The top-left inset shows the \grizli\ redshift probability function $p(z)$ in orange, with listed the best-fist \grizli\ spectroscopic redshift and upper and lower 3$\sigma$ values. The bottom-right insets show false-color images ($2^{\prime \prime} \times 2^{\prime \prime}$ from HST-WFC3/NIR bands (left) and JWST-NIRISS bands (right). {\bf Bottom:} Same as the top panel, but for HFF-DS 5607. {\it With a gravitational lensing corrected stellar mass below $10^{10}$~M$_{\odot}$, HFF-DS 5607 is the first low-mass quiescent galaxy spectroscopically confirmed at $z>2$.}}
 \label{fig:nirissspecs}
\end{figure*}

\section{Analysis and results} \label{sec:analysis}

The GLASS-ERS program, executed on June 28-29 2022, performed wide-field slitless spectroscopy at 1.0–2.2$\mu$m with NIRISS with a single pointing roughly centered on the core of A2744-clu. The reduction of the NIRISS data, along with the extraction of the spectra and modeling of the contamination, performed with the \grizli\footnote{\url{https://github.com/gbrammer/grizli}} analysis software \citep{Brammer2021}, is presented in another companion paper \citep[Paper I]{RobertsBorsani2022}.

Of the eight candidates listed in Table~\ref{tab:QGs}, three (IDs=1412, 2500, 4111) are outside of the NIRISS pointing, whereas three others (IDs=6555, 1103, an 5715) have a significant level of contamination from bright nearby sources which cannot be optimally corrected for with the current preliminary calibration files (see Paper I for details). The \grizli\ extracted 1D NIRISS spectra of the remaining two candidates are shown in Figure~\ref{fig:nirissspecs}, along with their false-color images constructed from both HST-WFC3/NIR bands and JWST-NIRISS bands separately. Although ID=4877 is the most massive quiescent candidate of the sample, with $\log{(M^{\rm corr}_{\rm star}/M_{\odot})}=10.67$ it is comparable to the lowest mass quiescent galaxy spectroscopically confirmed at $z>2$ to-date \citep{EstradaCarpenter2020}. On the other hand, ID=5607, with $\log{(M^{\rm corr}_{\rm star}/M_{\odot})}=9.92$, pushes the limit well below previous works, since it has a stellar mass estimate before lensing magnification correction $<10^{10.5}$~M$_{\odot}$.

Figure~\ref{fig:nirissspecs} shows the best-fit \grizli\ model (red curve) which places them, respectively, at redshifts of $z_{\rm spec}=2.650^{+0.004}_{-0.006}$ and $z_{\rm spec}=2.433^{+0.032}_{-0.016}$ (3$\sigma$ errors). The NIRISS spectra show strong Balmer/4000\AA~breaks, as well as other blended stellar continuum absorption features (the locations of CaII~HK, H$\delta$, H$\gamma$, H$\beta$, and NaD are highlighted in Fig.~\ref{fig:nirissspecs}). Furthermore, the NIRISS spectra do not show evidence of emission lines, confirming the quiescent nature of the candidates. For ID=4877, emission line forced photometry returns $f_{\rm [OII]}=0.2(\pm1.5)\times10^{-18}$~erg~s$^{-1}$~cm$^{-2}$, $f_{\rm H\beta}=1.4(\pm2.1)\times10^{-18}$~erg~s$^{-1}$~cm$^{-2}$, and $f_{\rm [OIII]}=4.6(\pm2.5)\times10^{-18}$~erg~s$^{-1}$~cm$^{-2}$, whereas for ID=5607 we measure $f_{\rm [OII]}=-4.7(\pm1.2)\times10^{-18}$~erg~s$^{-1}$~cm$^{-2}$, $f_{\rm H\beta}=-0.2(\pm2.3)\times10^{-18}$~erg~s$^{-1}$~cm$^{-2}$, although both [OII] and H$\beta$ are right at the longer wavelength end of the F115W and F150W spectra, respectively, whereas [OIII] falls in the gap between F150W and F200W. 

We used FAST to model the HFF-DeepSpace photometry fixing the redshift at the spectroscopic redshift. Compared to the SED-modeling assumptions presented in \S\ref{sec:sample}, we allowed for a range of stellar metallicities from sub-solar (Z=0.004) to super-solar (Z=0.05). Figure~\ref{fig:nirissspecs} shows in blue the best-fit FAST model. 

The best-fit stellar masses (not corrected for lensing magnification factor $\mu$) at the spectroscopic redshifts are $\log{(\mu \times M_{\rm star}/M_{\odot})}=10.84^{+0.17}_{-0.04}$ and $10.25^{+0.04}_{-0.06}$ for ID=4877 and ID=5607, respectively, in good agreement within the errors with those previously determined adopting the photometric redshifts. The SED modeling confirms the quiescent nature of these two candidates, with 1$\sigma$ upper limits on the star-formation rates of SFR$<$0.1-0.2~M$_{\odot}$~yr$^{-1}$, and on the sSFR of $\log{{\rm (sSFR[yr^{-1}])}}<-12$ and $<-11$, for ID=4877 and ID=5607, respectively, compared to $\log{({\rm sSFR[yr^{-1}]})}>-8.8$ for typical main-sequence star-forming galaxies at $2.0<z<2.5$ with similar stellar masses \citep{Whitaker2014}.

Using revised lensing magnification maps from \citet{Bergamini2022} and adopting the spectroscopic redshifts, we find $\mu_{\rm spec}=2.02\pm0.04$ and $\mu_{\rm spec}=2.09\pm0.05$ for ID=4877 and ID=5607, respectively. We note that this does not include systematic uncertainty arising from lens modeling assumptions, although we expect this uncertainty to be modest given the magnification factors.
The gravitational lensing magnification corrected stellar masses are therefore found to be $\log{(M^{\rm corr}_{\rm star}/M_{\odot})}=10.53^{+0.18}_{-0.06}$ for ID=4877 and $\log{(M^{\rm corr}_{\rm star}/M_{\odot})}=9.93^{+0.06}_{-0.07}$ for ID=5607. With a stellar mass below $10^{10}$~M$_{\odot}$, HFF-DS 5607 is the lowest mass quiescent galaxy to be spectroscopically confirmed at $z>2$ to-date.

\section{Summary and conclusions} \label{sec:conclusions}

This Letter presents a first look at low-mass quiescent galaxies in the JWST NIRISS data from the GLASS-ERS survey in the Abell 2744 cluster. The power of JWST for spectroscopic studies of low-mass quiescent galaxies at cosmic noon (i.e., $z>2$) is demonstrated by spectroscopic confirmation of the quiescent nature of two $z\sim2.5$ galaxies, which were identified through SED modeling as passive candidates in the HFF-DeepSpace photometric catalogs. Our results are summarized as follows: 

\begin{itemize}
    \item We measure the spectroscopic redshifts of two low-mass quiescent galaxies via their Balmer/4000\AA~breaks and stellar absorption features from NIRSS spectra ($z=2.650^{+0.004}_{-0.006}$ for HFF-DS 4877 and $z=2.433^{+0.032}_{-0.016}$ for HFF-DS 5607; 3$\sigma$ errors). 
    \item The quiescent nature of the two galaxies is further confirmed by SED modeling, resulting in SFR$<$0.2~M$_{\odot}$~yr$^{-1}$ and $\log{(sSFR[yr^{-1}])}<-11$, i.e., star formation activity suppressed by more than two orders of magnitude compared to typical star-forming galaxies with similar masses at $2.0<z<2.5$. 
    \item No nebular emission lines (i.e., [OII], H$\beta$, and [OIII]) are detected in the spectra, further supporting the quiescent nature of these two galaxies. 
    \item HFF-DS 5607, with a gravitational lensing magnification corrected stellar mass $\log{(M^{\rm corr}_{\rm star}/M_{\odot})}=9.93^{+0.06}_{-0.07}$, is the first spectroscopically confirmed low-mass quiescent galaxy at $z>2$.  
\end{itemize}

Soon-to-be-released updates and improvements to the existing NIRISS reference and calibration files are expected to further improve wavelength calibrations, spectral trace offset measurements, flux calibrations, and contamination modeling. This will enable for the confirmation of a larger sample of low-mass quiescent galaxies at $z>2$ in the Abell 2744 cluster ERS data, as well as allowing for the inclusion of the NIRISS spectra in the SED modeling for robust derivations of, e.g., SFH and stellar ages, thus providing the crucial information to discriminate between the multiple quenching mechanisms characterized by different timescales. JWST is finally allowing us, for the first time, to spectroscopically identify and study the population of low-mass ($\log{(M_{\rm star}/M_{\odot})}<10.5$) quiescent galaxies at $z>2$.

\begin{acknowledgments}
This work is based on observations made with the NASA/ESA/CSA James Webb Space Telescope. The data were obtained from the Mikulski Archive for Space Telescopes at the Space Telescope Science Institute, which is operated by the Association of Universities for Research in Astronomy, Inc., under NASA contract NAS 5-03127 for JWST. These observations are associated with program JWST-ERS-1324. We acknowledge financial support from NASA through grants JWST-ERS-1324. KG acknowledges support from Australian Research Council Laureate Fellowship FL180100060. We acknowledge financial support through grants PRIN-MIUR 2017WSCC32 and 2020SKSTHZ. AA has received funding from the European Union’s Horizon 2020 research and innovation programme under the Marie Skłodowska-Curie grant agreement No 101024195 — ROSEAU. MB acknowledges support from the Slovenian national research agency ARRS through grant N1-0238. 
\end{acknowledgments}

\bibliography{sample631}{}
\bibliographystyle{aasjournal}

\end{document}